\documentclass[runningheads]{llncs}

\usepackage{graphicx}
\usepackage{subcaption}
\usepackage{float}
\usepackage{placeins}
\usepackage{booktabs}
\usepackage{multirow}
\usepackage{amsmath}
\usepackage{microtype}
\usepackage{enumitem}
\usepackage{xcolor}
\usepackage{array}
\usepackage{tabularx}
\usepackage{tikz}
\usepackage[hidelinks,bookmarks=true,bookmarksnumbered=true]{hyperref}

\newcommand{\feat}[1]{{\small\texttt{\def\_{\textunderscore\allowbreak}#1}}}

\newcommand{\missingfig}[1]{%
  \fbox{\parbox{0.85\linewidth}{\centering\vspace{1.2cm}%
    \textit{#1}\vspace{1.2cm}}}}

\newcommand{\rulelabel}[5]{%
  \medskip
  \noindent\textbf{#1}\ifx&#2&\else~(\textit{#2})\fi:\par
  \noindent\hspace*{1.5em}#3\par
  \noindent\hspace*{3em}$\to$~#4\par
  \noindent\hspace*{1.5em}{\small\textit{#5}}\par
  \smallskip\noindent\ignorespaces
}

\begin{document}

\title{Interpretable Crisis Behavior Analysis Using Mobility and Social Media Data}
\titlerunning{Interpretable Crisis Behavior Analysis Using Mobility and Social Media Data}

\author{Muhammad Hamza Arshad Majeed, Sidahmed Benabderrahmane, Talal Rahwan}
\authorrunning{Majeed et al.}

\institute{New York University (NYUAD), Division of Science, Computer Science Department\\
\email{mm12283@nyu.edu}}

\maketitle

\begin{abstract}
Crises alter both how people move and how they communicate. During emergencies such as wildfires and pandemics, changes in mobility patterns and online emotional discourse evolve jointly, yet they are typically studied in isolation. This paper presents a unified and interpretable pipeline that integrates mobility and social media data to identify cross-domain behavioral patterns in crisis settings. The framework is evaluated through two case studies: a short-horizon analysis of the January 2025 Los Angeles wildfires (prototype case) and a longitudinal analysis of UAE COVID-19 behavior from March 2020 to December 2021 (primary case, 671 days). The pipeline aligns heterogeneous daily signals, transforms them into binary behavioral states, applies Formal Concept Analysis (FCA) to extract co-occurrence structure, mines association rules, and validates rule stability through chronological holdout testing. A structured policy-translation layer renders robust rules as operational briefs specifying triggers, lead times, and action playbooks. Results reveal clear cross-domain behavioral structure in both crises. In the wildfire case, traffic stress, fear/anger sentiment, and governance discourse are tightly coupled within a 33-day window, with key rules reaching 100\% confidence and lift scores up to 2.5. In the COVID case, repeated mobility adaptation and sentiment volatility yield 8 stable same-day rules (88\% holdout pass rate) and 40 clean predictive rules with 2--7 day lead horizons. The work demonstrates that interpretable multimodal fusion can produce both scientifically credible and policy-actionable crisis intelligence.

\textbf{Availability}: https://github.com/HamzaArshad2004/css-project-wildfire 

\end{abstract}

\section{Introduction}

Human behavior during crises is visible across both physical and digital spaces. People alter where they go, how frequently they travel, and how they express risk, uncertainty, trust, and frustration online. Studying only one of these channels yields an incomplete account of social response, and potentially misleads decision-makers who must act under uncertainty. This paper is motivated by the premise that crisis behavior should be modeled jointly across mobility and discourse. The practical rationale is direct: decision-makers need to know not only what the public is feeling, but whether those feelings are accompanied by real behavioral shifts. That distinction matters for communication timing, intervention targeting, and operational readiness. The paper addresses two qualitatively different crises to test whether behavioral coupling is a general phenomenon or an artifact of a single event type. The January 2025 Los Angeles wildfires represent an acute, geographically concentrated, rapidly evolving emergency. The UAE COVID-19 pandemic (2020--2021) represents a prolonged, policy-mediated, multi-wave disruption spanning 671 days and four distinct policy phases: initial lockdown (Mar--May 2020), phased reopening (Jun--Dec 2020), mass vaccination rollout (Jan--Aug 2021), and endemic transition (Sep--Dec 2021). Together, the two cases test whether the framework transfers across acute and prolonged crisis settings.

\paragraph{\textbf{Research Questions}:}
This work addresses four central questions:
\begin{enumerate}[noitemsep, topsep=2pt]
  \item How do mobility and online discourse signals co-evolve during crises of different types and durations?
  \item Which cross-domain behavioral patterns recur reliably within each crisis, and can they survive conservative temporal validation?
  \item Which patterns are contemporaneous, and which provide meaningful predictive lead time suitable for anticipatory intervention?
  \item Do behavioral patterns transfer across acute and prolonged crisis types?
\end{enumerate}

\paragraph{\textbf{Contributions}:}
The paper makes four primary contributions:
\begin{itemize}[noitemsep, topsep=2pt]
  \item An interpretable mobility--discourse fusion pipeline grounded in binary behavioral states and Formal Concept Analysis;
  \item A two-case evaluation covering a 33-day LA wildfire case (Caltrans PeMS + Reddit) and a 671-day UAE COVID-19 case (Google Mobility + Reddit);
  \item A reliability-first rule framework combining six-stage pruning, leakage control, and chronological holdout testing;
  \item A structured policy-brief format translating validated rules into agency-ready operational guidance with named bodies, lead times, and action playbooks.
\end{itemize}

\section{Related Work}

\paragraph{\textbf{Crisis mobility}.}
Prior work has shown that large-scale crises strongly reshape human movement. Beria and Lunkar~\cite{beria2021} used anonymized location-based services data to study mobility restrictions during the first COVID-19 wave in Italy, while Caselli et al.~\cite{caselli2020} showed that post-lockdown mobility recovery was shaped by policy and socioeconomic factors and could be partly forecast from early crisis data. These studies establish mobility as a key behavioral signal, but they largely treat movement as a standalone outcome rather than modeling its relationship with online discourse. This paper addresses this gap by jointly analyzing mobility and discourse at daily resolution.

\paragraph{\textbf{Social media as a behavioral sensor}.}
Social media has also been widely used as a near-real-time sensor of public response during emergencies. Imran et al.~\cite{imran2015} surveyed machine learning pipelines for crisis-related social media processing, Zhang et al.~\cite{zhang2019} reviewed disaster-oriented public information systems, and Zubiaga et al.~\cite{zubiaga2018} studied rumor detection in noisy, fast-evolving crisis content. For sentiment quantification, this paper uses VADER~\cite{hutto2014}, a rule-based model designed for social media text and suitable when labeled crisis-specific data are unavailable. However, these works typically analyze online discourse separately from physical mobility.

\paragraph{\textbf{Integration gap}.}
Prior studies establish mobility and social discourse as valuable but mostly separate crisis signals. Existing multimodal approaches often favor predictive accuracy through latent embeddings or black-box classifiers, limiting auditability for emergency practitioners. This paper addresses this gap with a transparent rule-based fusion framework that links mobility, sentiment, emotion, and topic states, validates the resulting patterns temporally, and translates robust rules into operational guidance.

\section{Data and Method}

\paragraph{\textbf{Case studies and data sources}.}
We evaluate the framework on two crises. The prototype case examines January 2025 wildfire behavior in Los Angeles over a 33-day window (January 3--February 4, 2025), covering four phases: pre-ignition baseline (Jan 3--6), rapid ignition and spread of the Palisades and Eaton fires (Jan 7--11), peak crisis with approximately 180,000 residents evacuated (Jan 12--20), and progressive containment and return orders (Jan 21--Feb 4). Mobility signals are daily Caltrans PeMS traffic metrics: Vehicle Miles Traveled (VMT), Vehicle Hours Traveled (VHT), Travel Time Index (TTI), and vehicle delay. Social discourse is collected with PRAW Python API from \texttt{r/LosAngeles} and \texttt{r/wildfires}, then processed with VADER sentiment scoring and keyword-based emotion detection. The primary case covers UAE COVID-19 behavior from March 1, 2020 to December 31, 2021 (671 days), combining national-level Google Community Mobility Reports with COVID-related discourse from \texttt{r/UAE}, \texttt{r/dubai}, \texttt{r/abudhabi}, \texttt{r/Coronavirus}, and \texttt{r/worldnews}. The UAE timeline is divided into four policy phases: initial lockdown (Mar--May 2020), phased reopening (Jun--Dec 2020), vaccination rollout (Jan--Aug 2021), and endemic transition (Sep--Dec 2021). In both cases, mobility and discourse signals are synchronized to a daily index; Reddit posts are deduplicated by post ID, and days with fewer than 5 posts are flagged as low-coverage. 

\paragraph{\textbf{Behavioral-state encoding}.}
The core design choice is to represent crisis dynamics as interpretable binary behavioral states. Each observation day is encoded as a set of semantically meaningful attributes, enabling transparent rule extraction, hierarchical concept analysis, and policy-legible interpretation. Feature engineering produces a 19-column LA matrix and a 29-column UAE matrix covering five groups of states: \textit{mobility states}, such as retail/workplace/transit drops, residential increase, grocery spike, and traffic congestion; \textit{sentiment states}, such as high negative sentiment, high positive sentiment, and day-over-day sentiment shift; \textit{emotion/topic states}, such as fear, anger, sadness, solidarity, health concern, vaccine mention, and \textit{policy/governance discussion}; composite states, such as \feat{emotion\_with\_mobility\_signal}, \feat{emotion\_mobility\_mismatch}, and \feat{calm\_mobile\_baseline}; and \textit{temporal states}, using lagged antecedents at 1, 3, and 7 days and lead outcomes at 2 and 7 days. Continuous signals are binarized using documented thresholds, including VADER polarity cutoffs, mobility deviations from baseline, and rolling Caltrans congestion baselines; full threshold definitions are provided in the repository. UAE thresholds are relative to the Google Mobility pre-pandemic baseline, while LA thresholds use a 30-day rolling pre-fire Caltrans average. The feature \feat{dominant\_emotion\_fear} is active when fear keywords exceed all other emotion categories on a given day.

\begin{figure}[t]
\centering
\scriptsize

\begin{minipage}[t]{0.36\linewidth}
\centering
\textbf{Formal context}\\[2pt]
\begin{tabular}{lccccccc}
\hline
Day & T & A & F & P & S & M & C \\
\hline
$d_1$ & 1 & 1 & 1 & 1 & 0 & 0 & 0 \\
$d_2$ & 1 & 0 & 1 & 1 & 1 & 0 & 0 \\
$d_3$ & 1 & 1 & 1 & 0 & 0 & 0 & 0 \\
$d_4$ & 0 & 0 & 1 & 1 & 0 & 1 & 0 \\
$d_5$ & 0 & 0 & 0 & 0 & 0 & 0 & 1 \\
\hline
\end{tabular}

\vspace{2pt}
\raggedright
\emph{T}: traffic; \emph{A}: anger; \emph{F}: fear;
\emph{P}: policy; \emph{S}: sadness;
\emph{M}: mismatch; \emph{C}: calm.
\end{minipage}
\hfill
\begin{minipage}[t]{0.27\linewidth}
\centering
\textbf{Lattice excerpt}\\[2pt]
\begin{tikzpicture}[
  scale=0.68,
  every node/.style={circle, draw, inner sep=1.1pt, font=\tiny},
  every path/.style={line width=0.3pt}
]
\node (top) at (0,2.3) {$\emptyset$};
\node (F) at (0,1.55) {$F$};
\node (TF) at (-0.9,0.75) {$T,F$};
\node (FP) at (0.9,0.75) {$F,P$};
\node (TAF) at (-1.3,-0.05) {$T,A,F$};
\node (TFP) at (0,-0.05) {$T,F,P$};
\node (FPM) at (1.3,-0.05) {$F,P,M$};
\node (C) at (1.65,1.55) {$C$};

\draw (top)--(F);
\draw (top)--(C);
\draw (F)--(TF);
\draw (F)--(FP);
\draw (TF)--(TAF);
\draw (TF)--(TFP);
\draw (FP)--(TFP);
\draw (FP)--(FPM);
\end{tikzpicture}
\end{minipage}
\hfill
\begin{minipage}[t]{0.33\linewidth}
\centering
\textbf{Example rules}\\[2pt]
\begin{tabular}{p{0.70\linewidth}c}
\hline
Rule & (sup., conf., lift) \\
\hline
$T+A \Rightarrow F$ & (40, 100, 1.25) \\
$A \Rightarrow T+F$ & (40, 100, 1.67) \\
$S \Rightarrow T+F+P$ & (20, 100, 2.50) \\
$M \Rightarrow F+P$ & (20, 100, 1.67) \\
\hline
\end{tabular}
\end{minipage}

\caption{Illustrative FCA example for crisis behavior analysis. The formal context encodes daily mobility--discourse states; the lattice excerpt shows selected shared behavioral patterns; and the rules summarize interpretable co-occurrences between traffic, emotion, policy discourse, mismatch, and calm states. Values denote support, confidence, and lift.}
\label{fig:fca_toy_example}
\vspace{-2em}
\end{figure}
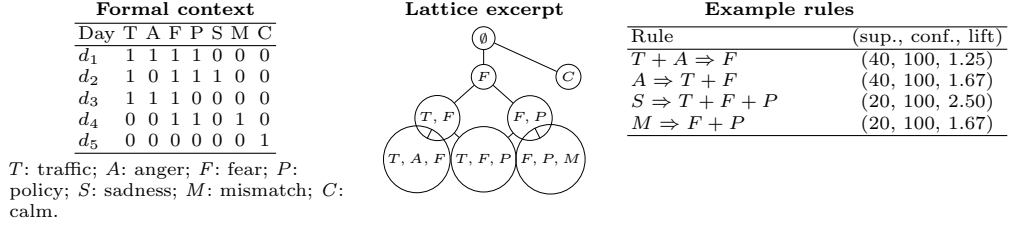

\paragraph{\textbf{Interpretable FCA and rule-mining layer}.}
The interpretable modeling layer is based on Formal Concept Analysis (FCA)~\cite{ganter1999} and association rule mining~\cite{agrawal1994}. Each crisis day is represented as a formal context, where objects are observation days and attributes are binary mobility, sentiment, emotion, topic, composite, or temporal states. FCA organizes this binary context into a concept lattice, where each concept corresponds to a set of days sharing a set of behavioral attributes. To make this step concrete, Fig.~\ref{fig:fca_toy_example} gives a small wildfire-like example using the same type of daily crisis features as our data: traffic congestion, anger, fear, policy discourse, sadness, mismatch, and calm baseline. The figure shows how daily binary observations are transformed into a formal context, how selected shared patterns appear in a lattice excerpt, and how association rules summarize interpretable behavioral regularities. This makes the resulting patterns auditable: for example, days combining traffic congestion and anger can be directly inspected and linked to fear, policy discourse, or mobility disruption. Temporal lag/lead features extend this layer from same-day pattern discovery to short-horizon prediction, while chronological validation checks whether the rules persist beyond the phase in which they were discovered.

\paragraph{\textbf{FCA construction and rule extraction}.}
For the UAE case, objects are aggregated by weekly majority vote for lattice construction to avoid combinatorial explosion, while rule mining runs on the full 671-day daily data. Same-day association rules are extracted using fixed thresholds: UAE rules require support $\geq10\%$, confidence $\geq75\%$, and lift $\geq1.8$; LA rules require support $\geq10\%$, confidence $\geq80\%$, lift $\geq1.05$, maximum premise size $=2$, and maximum conclusion prevalence $\leq75\%$. A tautology filter removes rules whose conclusions follow definitionally from composite-feature premises, and rules are tagged \texttt{cross\_domain = True} when they span both mobility and emotion/discourse feature sets.

\paragraph{\textbf{Temporal prediction, pruning, and validation}.}
To identify predictive structure, lag and lead features are generated by shifting signals across 1-, 3-, and 7-day antecedent windows and 2- and 7-day consequent windows, yielding a $671{\times}111$ UAE feature matrix. This matrix supports rules where current signals predict outcomes several days later. A six-stage filtering pipeline reduces 3,483 raw predictive candidates to 40 clean rules by removing same-lag artifacts, same-day contamination, \texttt{\_lead} features in antecedents, tautological feature-group rules, bidirectional duplicates, and rules subsumed by stronger antecedents. Chronological validation uses a 70/30 split for the UAE case (training: Mar 2020--Jun 2021; holdout: Jul 2021--Dec 2021), retaining only rules whose confidence drops by at most 10 percentage points in the holdout period. GPT-4o-mini is used only after statistical filtering as an evaluation layer, scoring each candidate for novelty and policy relevance on a 1--10 scale. Novelty scores of 9--10 are reserved for rules where an emotion/discourse signal today forecasts a mobility change 2--7 days later. The system prompt includes the UAE policy timeline and requires named-agency recommendations (NCEMA, DHA, MoHAP, WAM for UAE; LAFD, CAL FIRE, LA County OES for LA).

\paragraph{\textbf{Policy-brief generation}.}
Each validated rule is translated into a structured operational brief specifying trigger conditions, expected outcome, statistical strength (support, confidence, lift, conviction, leverage), signal window, agency action playbook, false-alarm risk level, and confirmation guidance. Predictive briefs additionally report \texttt{lead\_time\_days} and a named responsible agency.

\section{Results}

\subsection{Case Study I: LA Wildfires}
\subsubsection{Behavioral activation profile.} Over the 33-day wildfire window (19 features), the most active binary states were: solidarity messages (54.5\%), policy/governance discussion (48.5\%), traffic congestion (42.4\%), fear keywords (39.4\%), sentiment shift (39.4\%), high negative sentiment (33.3\%), and sadness keywords (33.3\%). Both \feat{emotion\_with\_mobility\_signal} and \feat{emotion\_mobility\_mismatch} activated on 27.3\% of days, indicating that emotional escalation sometimes translated into observable mobility shifts (peak crisis, Jan 12--20) and sometimes did not (containment phase, when fear persisted but congestion normalized).

\subsubsection{Top cross-domain wildfire rules.} Five representative high-value cross-domain rules are summarized in Table~\ref{tab:wildfire_rules}; the three most important are described below.

\begin{table}
\vspace{-2 em}

\tiny
\centering
\caption{Top Cross-Domain Association Rules: LA Wildfires}
\label{tab:wildfire_rules}
\begin{tabularx}{\linewidth}{@{}>{\raggedright\arraybackslash}X c c c@{}}
\toprule
\textbf{Antecedent $\to$ Consequent} & \textbf{Supp.} & \textbf{Conf.} & \textbf{Lift} \\
\midrule
traffic congestion + anger $\to$ fear keywords       & 15.2\% & 100\%  & 2.54 \\
traffic congestion + sadness $\to$ policy/governance & 18.2\% & 83.3\% & 1.72 \\
traffic congestion + anger $\to$ policy/governance   & 15.2\% & 80.0\% & 1.65 \\
fear + anger $\to$ traffic congestion                & 18.2\% & 100\%  & 2.30 \\
anger + policy/governance $\to$ traffic congestion   & 15.2\% & 100\%  & 2.30 \\
\bottomrule
\end{tabularx}
\vspace{-2.5 em}

\end{table}

W1: \feat{traffic\_congestion}+\feat{anger}
$\Rightarrow$
\feat{fear\_keywords}. 
%
On every day that combined traffic congestion with anger discourse, fear keywords also appeared. This 100\% confidence rule suggests that physical immobility and anger toward conditions reliably co-occur with the escalation to explicit fear expression, a pattern consistent with acute, place-based threat.

W2: \feat{traffic\_congestion}+\feat{sadness} $\Rightarrow$ \feat{policy/governance}. 
On days where congestion co-occurred with grief and loss discourse, public accountability framing consistently emerged. Governance communication quality is most urgently needed precisely when communities experience both physical disruption and emotional loss simultaneously.

W3: fear+anger $\Rightarrow$ traffic congestion. The reverse coupling reached lift 2.30, indicating that a fear--anger climate was more than twice as likely as chance to co-occur with detectable traffic stress. This is the most operationally relevant direction: discourse signals offer a potential early-warning indicator for congestion-management agencies.

\begin{figure}[htbp]
\centering

\begin{minipage}[t]{0.46\linewidth}
\centering
\IfFileExists{results/visualizations/la_wildfire_mobility_sentiment.png}{%
    \includegraphics[width=\linewidth]{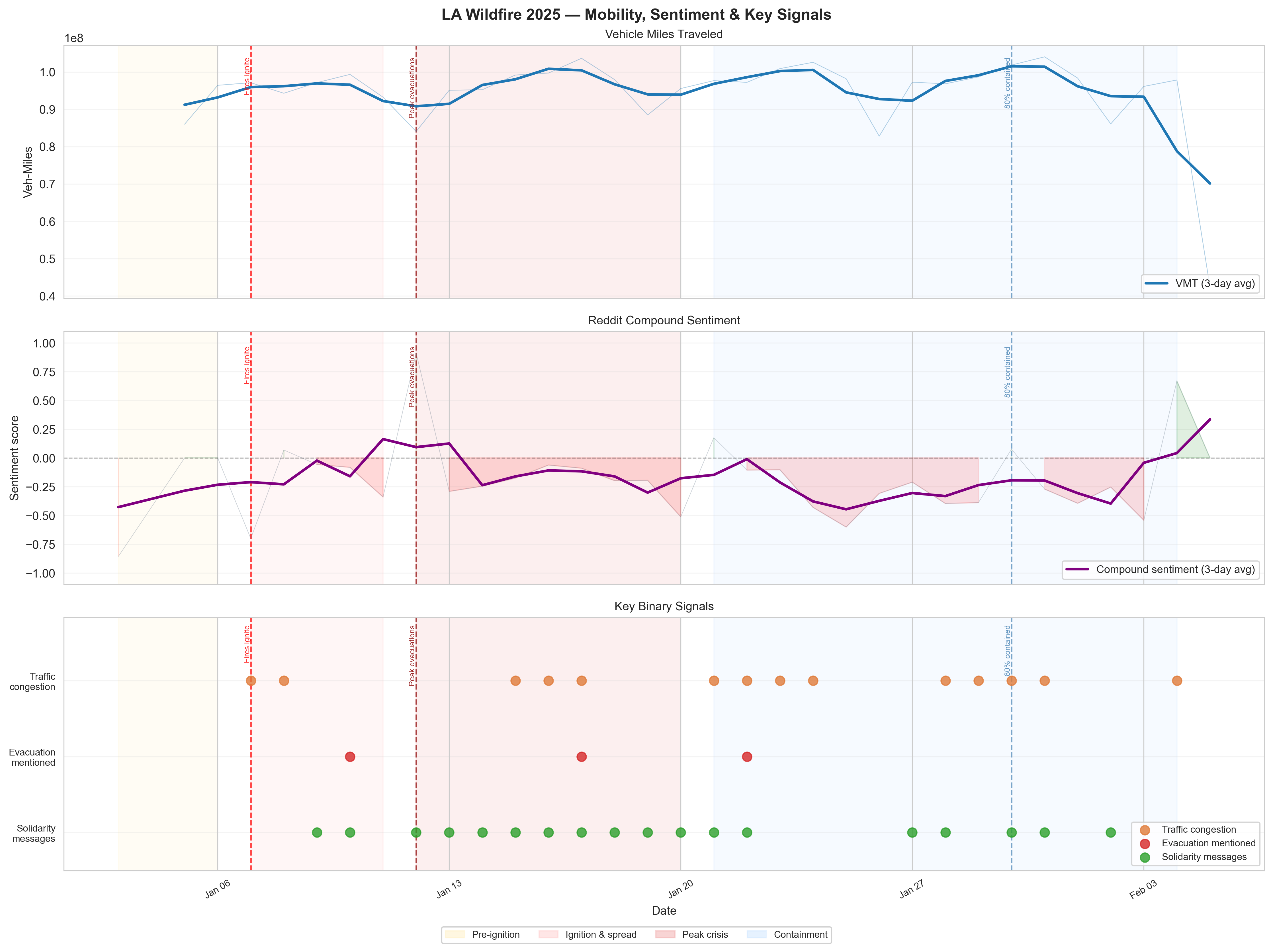}%
}{%
    \missingfig{LA wildfire mobility and sentiment co-evolution.}%
}
\end{minipage}
\hfill
\begin{minipage}[t]{0.46\linewidth}
\centering
\IfFileExists{results/visualizations/la_wildfire_rules.png}{%
    \includegraphics[width=\linewidth]{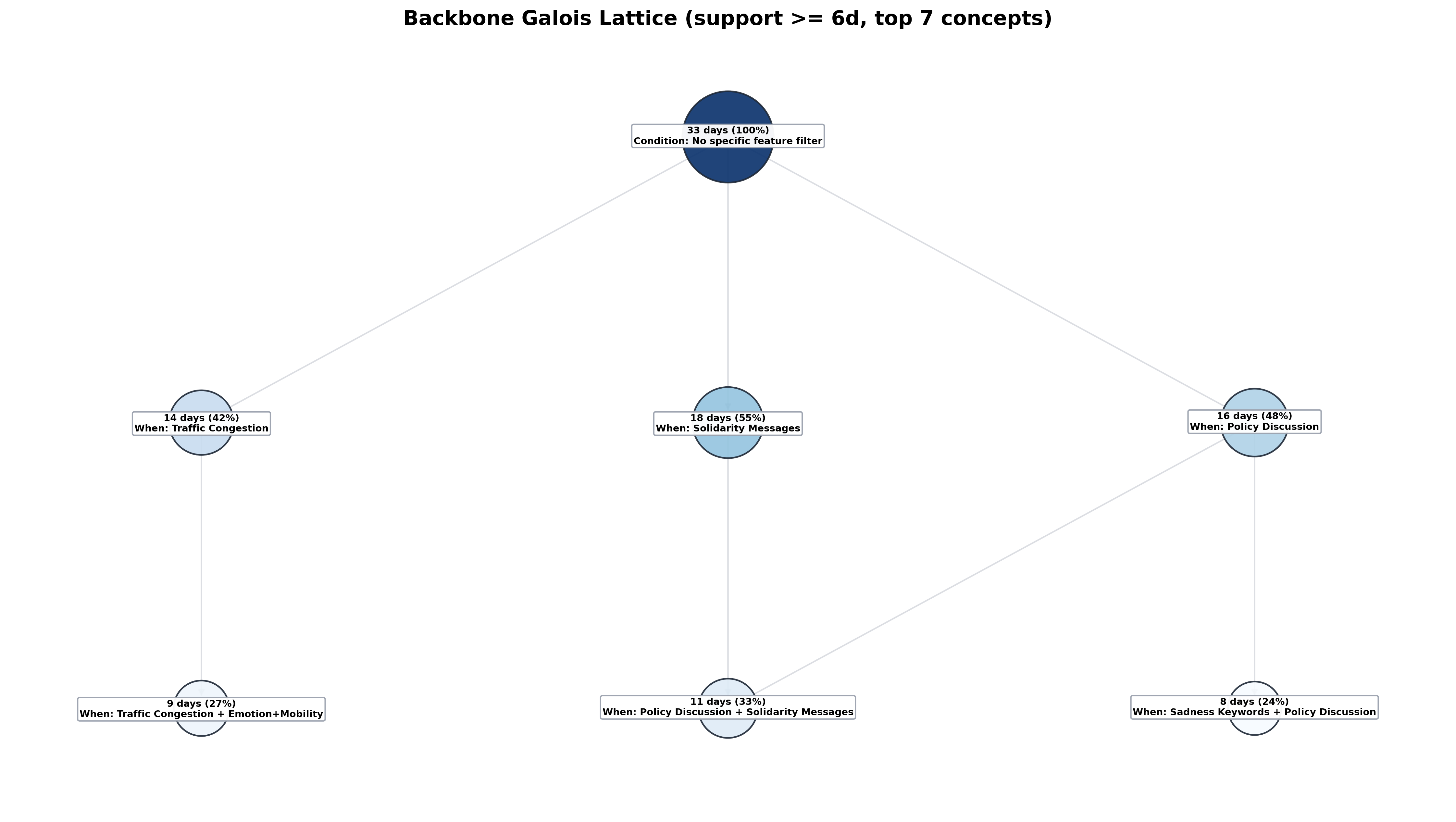}%
}{%
    \missingfig{LA wildfire association rules or FCA lattice.}%
}
\end{minipage}

\caption{LA wildfire analysis. Left: co-evolution of Caltrans mobility indicators and Reddit discourse during the January--February 2025 wildfire event. Right: FCA concept lattice extracted from the behavioral-state representation.}
\label{fig:la_wildfire_combined}
\vspace{-3.5 em}

\end{figure}

\FloatBarrier

\subsection{Case Study II: UAE COVID-19}
\subsubsection{Behavioral activation profile.} 
\begin{figure}[htbp]
  \centering
  \begin{minipage}[t]{0.41\linewidth}
    \centering
    \includegraphics[width=\linewidth]{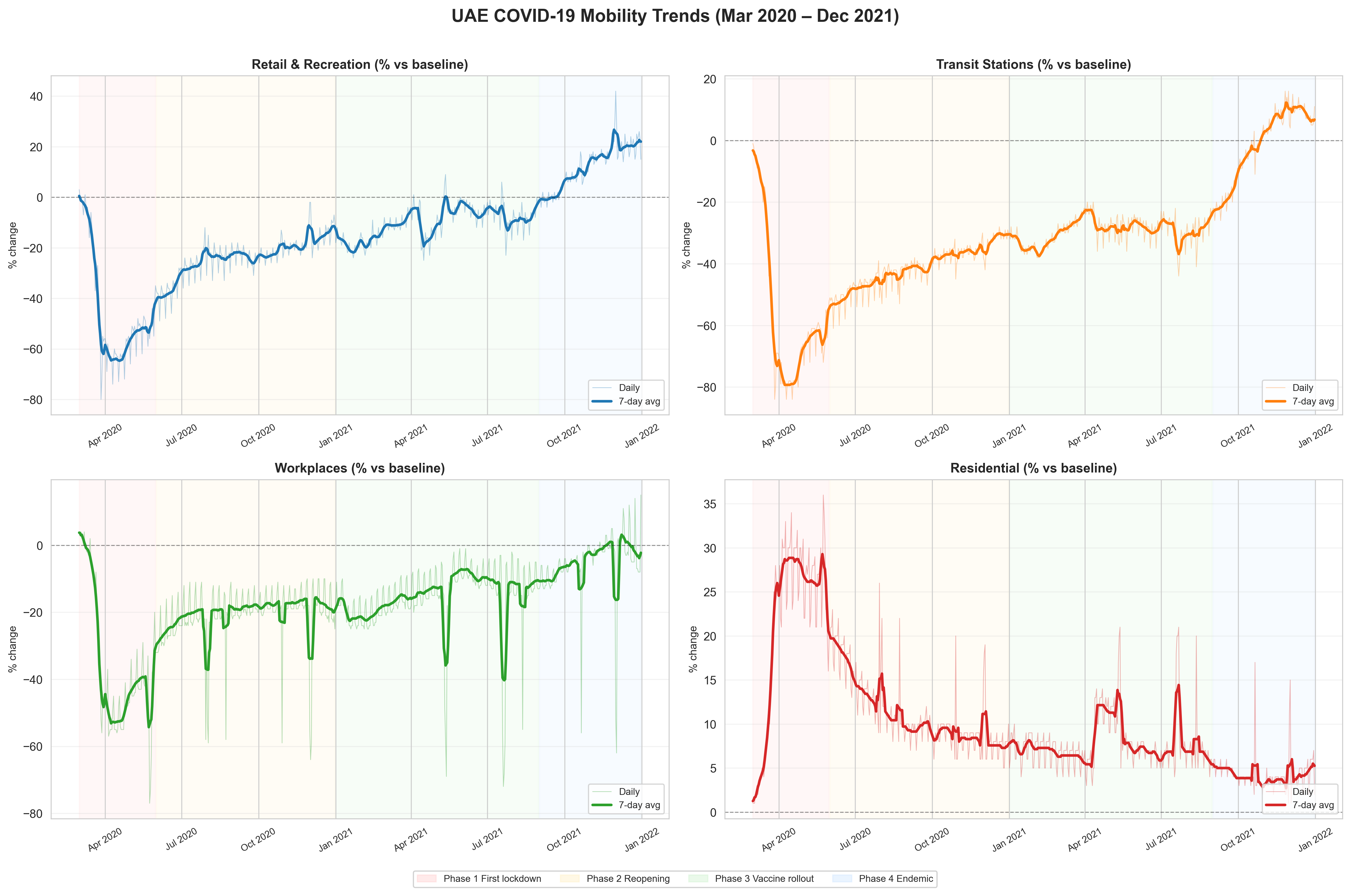}
    \smallskip
    {\small (a) UAE mobility trends with 7-day rolling average and policy-phase shading.}
  \end{minipage}
  \hfill
  \begin{minipage}[t]{0.41\linewidth}
    \centering
    \includegraphics[width=\linewidth]{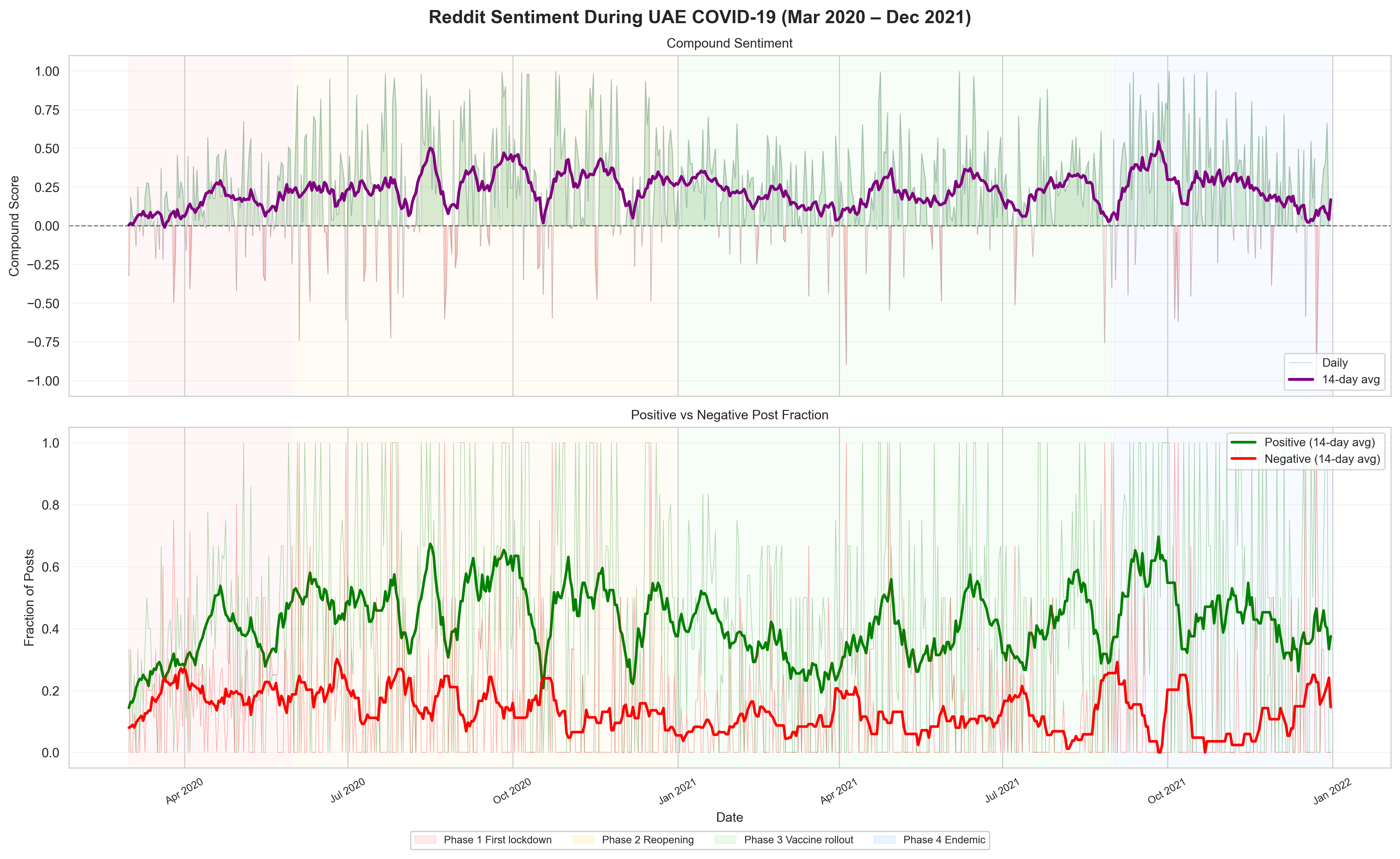}
    \smallskip
    {\small (b) VADER compound score and positive/negative fractions with 14-day rolling average.}
  \end{minipage}
  \caption{Macro temporal dynamics of movement and discourse during UAE COVID-19. Sharp mobility drops in Phase 1 and partial recovery in Phases 3--4 are visible alongside the persistent fear-dominant sentiment baseline.}
  \label{fig:macro_dynamics}
  \vspace{-3 em}

\end{figure}

Over the 671-day COVID window (29 features), major binary-state activation rates were:
\begin{itemize}[noitemsep, topsep=3pt, leftmargin=1.2em]
  \item \feat{covid\_topic\_detected}, \feat{dominant\_emotion\_fear}: 84.9\%
  \item \feat{sentiment\_shift\_detected}: 58.4\%; \feat{health\_concern}: 58.3\%
  \item \feat{mobility\_drop\_transit}: 54.1\%; \feat{vaccine\_mentioned}: 45.0\%
  \item \feat{solidarity\_messages}: 44.0\%; \feat{residential\_increase}: 43.1\%
  \item \feat{grocery\_spike}: 42.6\%; \feat{calm\_mobile\_baseline}: 39.6\%
\end{itemize}
This profile reflects sustained concern and iterative adaptation rather than a single behavioral shock, consistent with the multi-wave pandemic structure. The macro-level temporal dynamics of mobility and discourse are shown in Figure~\ref{fig:macro_dynamics}.

\subsubsection{Rule quality outcomes.} Both pipelines applied conservative filtering: 23 raw same-day rules extracted; 9 after pruning; \textbf{8 stable after chronological holdout (88\%)}; and \textbf{40 clean predictive rules} after six-stage pruning. The one rule that failed, \feat{grocery\_spike} + \feat{weekend} $\to$ \feat{calm\_mobile\_baseline} showed a 12.2 pp confidence drop on the holdout period, consistent with the gradual erosion of weekend-specific behavioral rhythms during Phase 4.

\subsubsection{\textbf{Top stable same-day COVID rules:}}
\paragraph{}
~~C1: \feat{health\_concern}+\feat{calm\_mobile\_baseline} $\Rightarrow$ \feat{grocery\_spike} (sup.=20.4\%, conf.=77.4\%, lift=1.82, nov.=8/10, pol.=7/10)

C2: \feat{grocery\_spike}+\feat{high\_pos\_sentiment} $\Rightarrow$ \feat{calm\_mobile\_baseline} (sup.=15.4\%, conf.=85.4\%, lift=2.16, nov.=7/10, pol.=6/10)

C3: \feat{grocery\_spike}+\feat{vaccine\_mentioned} $\Rightarrow$ \feat{calm\_mobile\_baseline} (sup.=26.2\%, conf.=79.5\%, lift=2.01, nov.=6/10, pol.=5/10)

Together, these rules describe the transition from latent health concern to behavioral normalization. C1 indicates that grocery-demand surges can emerge even on otherwise routine mobility days when health concerns remain elevated. In contrast, C2 and C3 associate grocery activity with positive sentiment and vaccine discussion, suggesting that later-period retail activity increasingly reflected adaptation and recovery rather than anxiety-driven stockpiling. The persistence of C3 across 26\% of all days highlights the stability of this normalization phase.

\begin{figure}[t!]
  \centering
  \begin{minipage}[t]{0.5\linewidth}
    \centering
    \includegraphics[width=\linewidth]{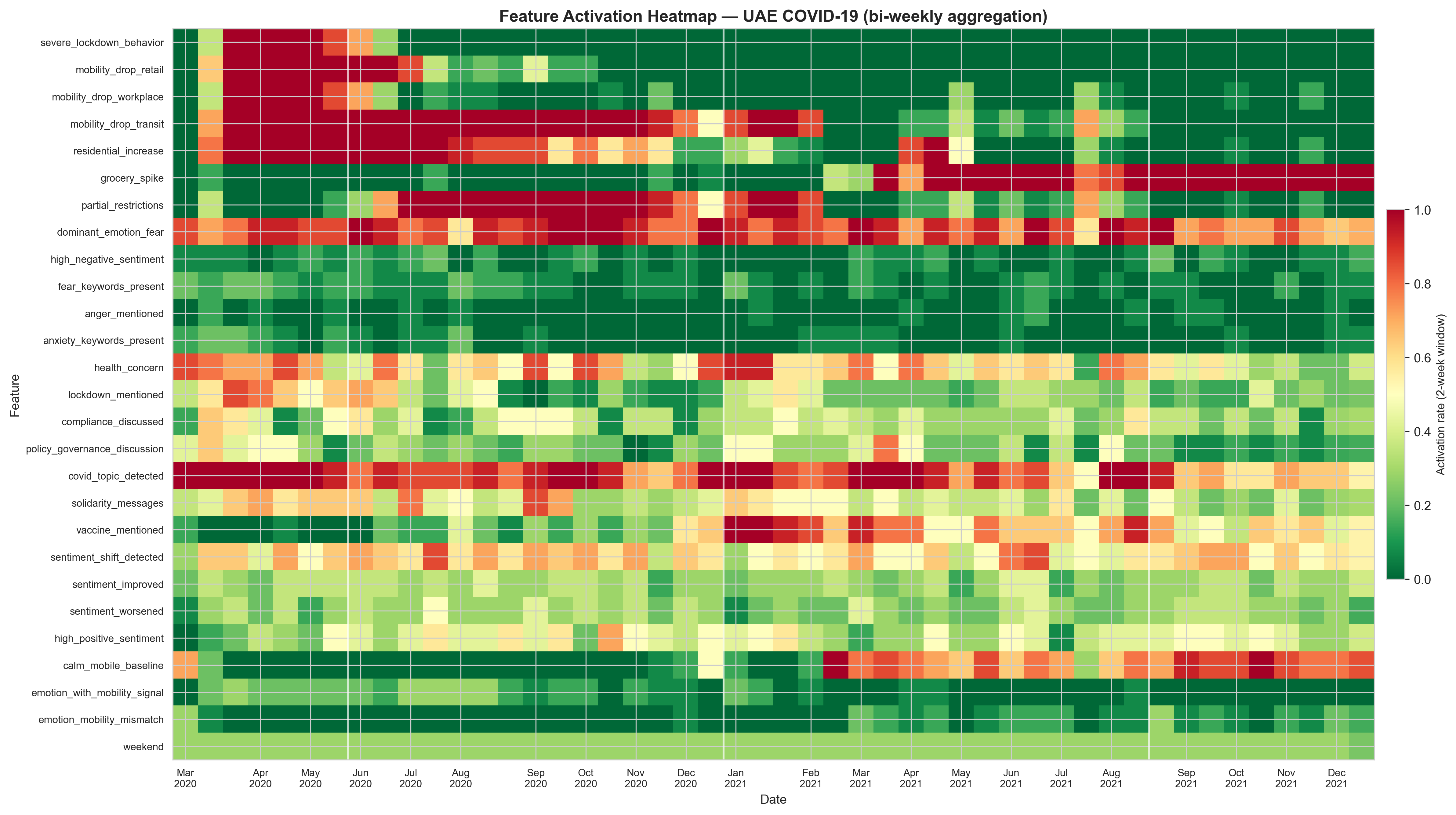}
    \smallskip
    {\small (a) Bi-weekly feature activation heatmap. Darker cells indicate higher activation.}
  \end{minipage}
  \hfill
  \begin{minipage}[t]{0.5\linewidth}
    \centering
    \includegraphics[width=\linewidth]{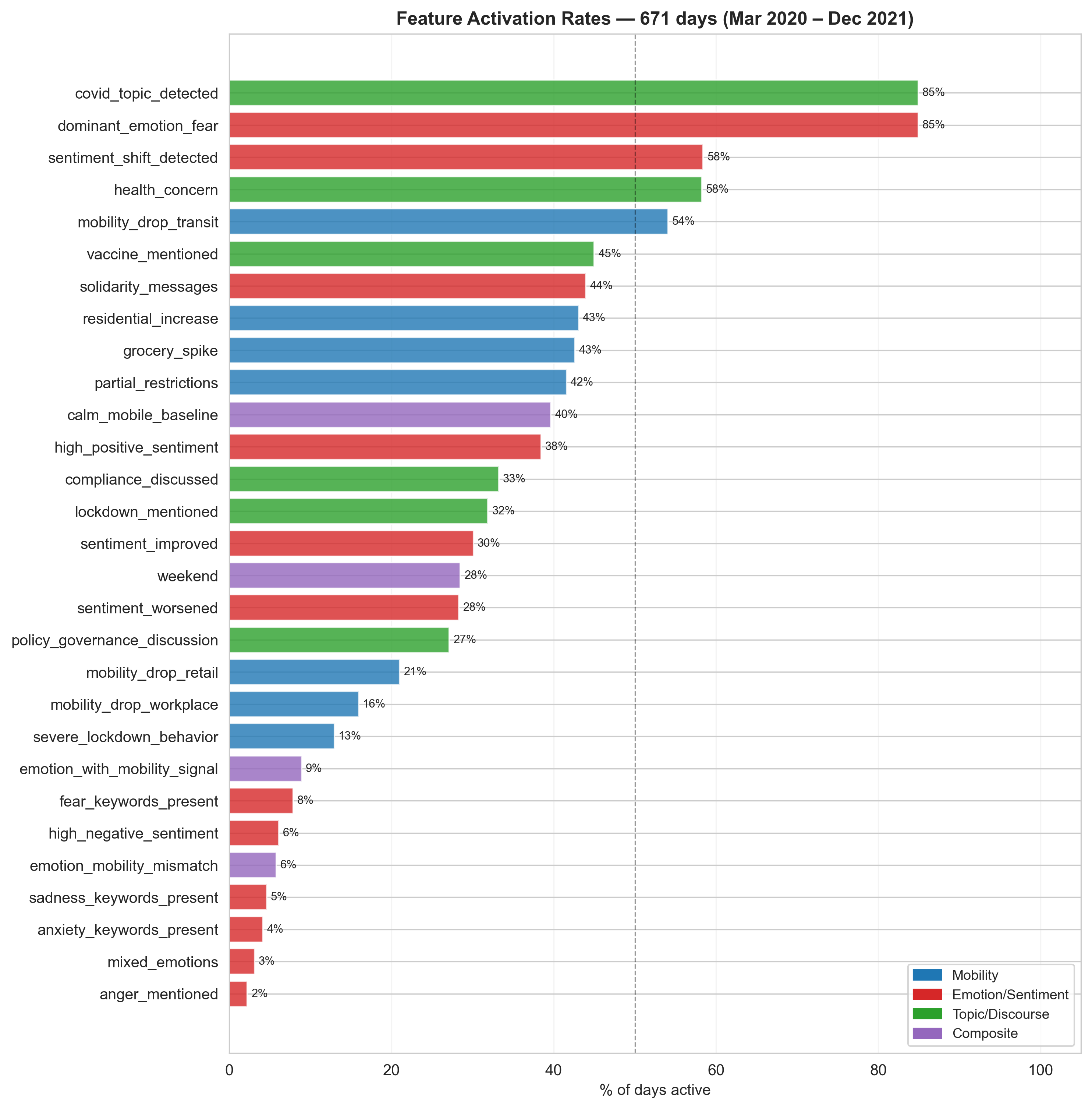}
    \smallskip
    {\small (b) Global activation rates of all 29 features, color-coded by domain.}
  \end{minipage}
  \caption{Behavioral-state structure and distribution in the UAE COVID case. The heatmap reveals phase-specific activation clusters; the bar chart shows that fear and COVID-topic features dominate across the full 671-day period, while grocery and calm-baseline features mark recovery windows.}
  \label{fig:feature_structure}
  \vspace{-1.8 em}
\end{figure}
\subsubsection{Top predictive COVID rules (lag/lead):}
\paragraph{}
~~~P1: grocery spike + lockdown mention $\Rightarrow$ vaccine mention \emph{(7-day lead)} (10.4, 85.3, 1.90, 9, 8)

P2: grocery spike + \feat{high\_pos\_sentiment\_lag1} $\Rightarrow$ calm mobile baseline \emph{(7-day lead)} (16.0, 81.0, 2.09, 8, 8)

P3: positive sentiment + calm mobile baseline $\Rightarrow$ grocery spike \emph{(2-day lead)} (14.9, 88.8, 2.08, 7, 7)

Together, these predictive rules show that discourse and consumer activity provide short-horizon behavioral signals. Lockdown discourse with grocery spikes precedes vaccine discussion by one week (P1), while positive sentiment combined with grocery activity forecasts stable mobility (P2). Conversely, stable positive days precede grocery upticks within 48 hours (P3), making these rules useful for communication timing and logistics pre-positioning. Values denote support, confidence, lift, novelty, and policy relevance.
\subsection{Cross-Case Comparison}
Table~\ref{tab:cross_case} summarizes the structural differences between the two case studies. Together, both cases confirm that coupling between discourse and mobility is real but conditional on crisis type, phase, and feature combination. Days of mismatch, where emotional and mobility signals diverge, are analytically meaningful rather than noise.

\begin{figure*}[t!]

  \centering
  \begin{subfigure}[t]{0.48\textwidth}
    \centering
    \includegraphics[width=\linewidth]{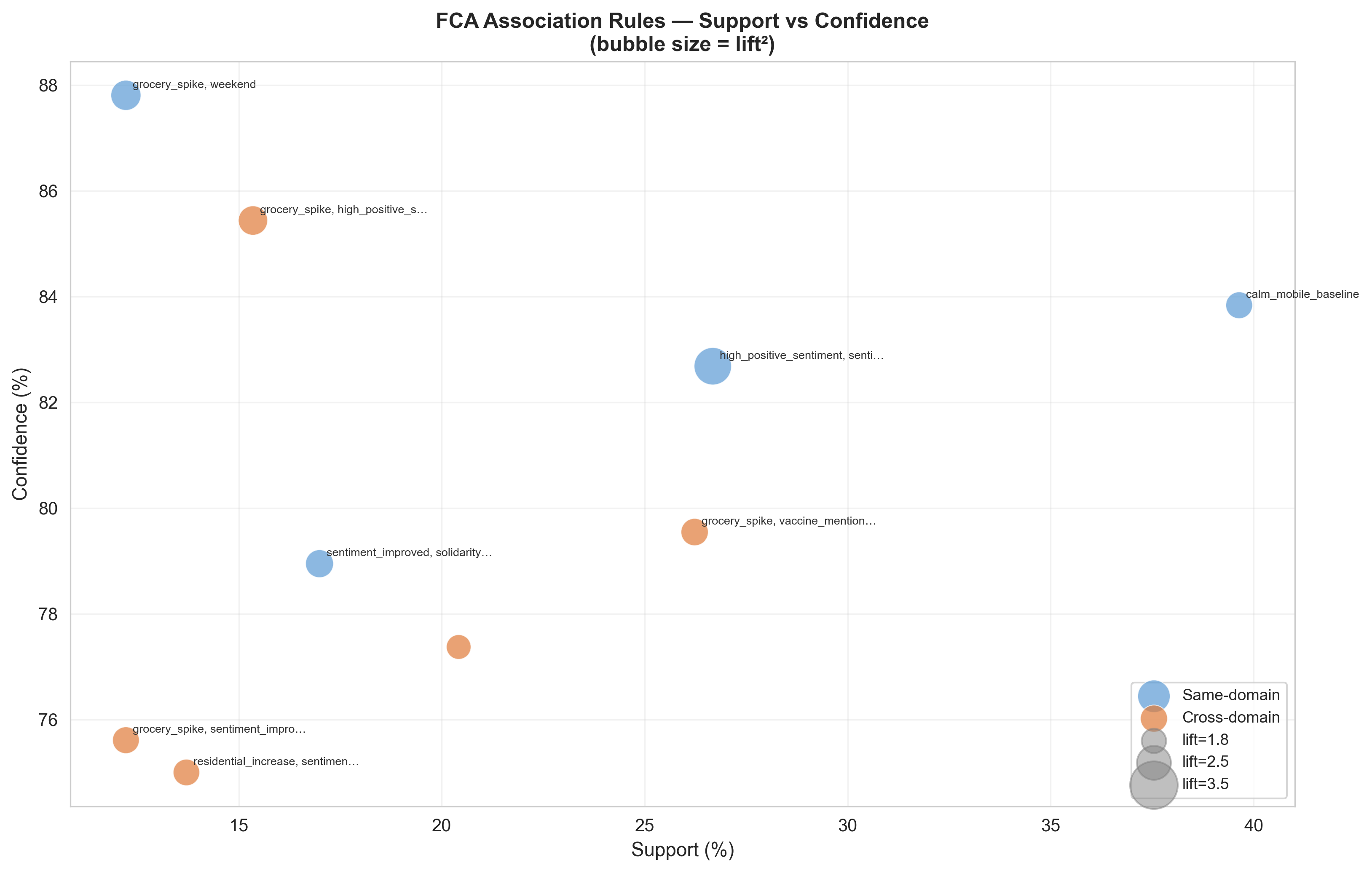}
    \caption{Rule quality: support vs.\ confidence bubble chart; bubble area $\propto$ lift$^2$; color = cross-domain vs.\ intra-domain.}
    \label{fig:rules_overview}
  \end{subfigure}
  \hfill
  \begin{subfigure}[t]{0.48\textwidth}
    \centering
    \includegraphics[width=\linewidth]{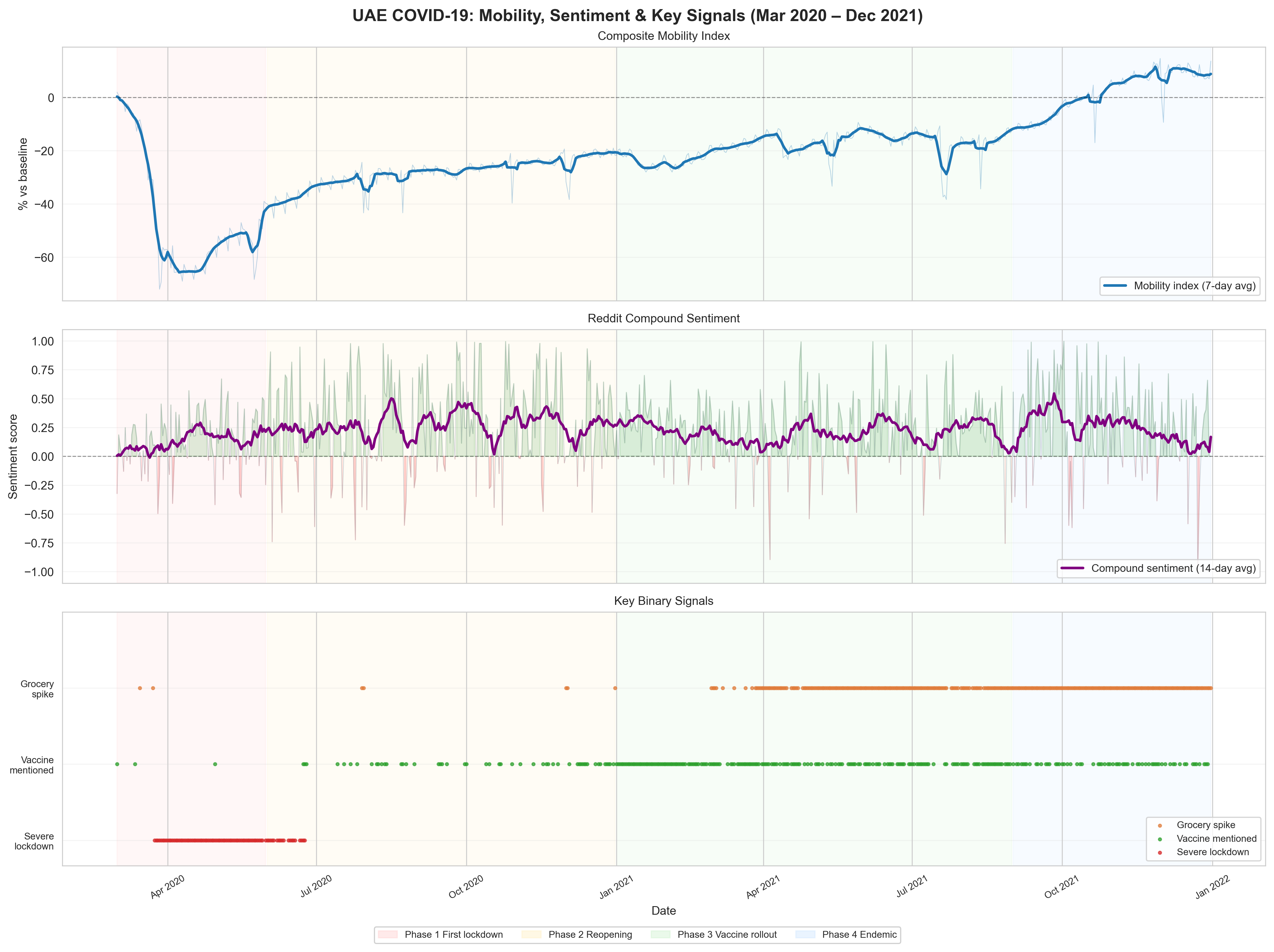}
    \caption{Three-panel view: mobility index / VADER sentiment / binary key signals, with policy-phase band overlay.}
    \label{fig:combined_analysis}
  \end{subfigure}
  \caption{Cross-domain rule landscape and integrated interpretation view. Cross-domain rules cluster at higher lift values than intra-domain rules, confirming that mobility--discourse coupling carries more statistical information than within-domain associations alone.}
  \label{fig:rule_and_combined}
  \vspace{-1 em}
\end{figure*}

\begin{table}[t!]
\vspace{-1 em}

\tiny
\centering
\caption{Comparison of Wildfire and COVID Case Study Outcomes}
\label{tab:cross_case}
\begin{tabularx}{\linewidth}{@{}>{\raggedright\arraybackslash}p{0.30\linewidth} >{\raggedright\arraybackslash}X >{\raggedright\arraybackslash}X@{}}
\toprule
\textbf{Dimension} & \textbf{LA Wildfire} & \textbf{UAE COVID-19} \\
\midrule
Observation days    & 33       & 671 \\
Binary features     & 19       & 29 \\
Raw same-day rules  & $\sim$30 & 23 \\
After pruning       & 5 (top)  & 9 \\
Holdout reliability & Prototype only & 88\% (8/9) \\
Predictive rules    & Limited  & 40 clean \\
Dominant discourse  & Governance + solidarity & COVID salience + fear \\
Dominant mobility   & Traffic congestion (TTI) & Transit suppression + grocery cycles \\
Mobility source     & Caltrans PeMS & Google Mobility Reports \\
Discourse source    & \texttt{r/LosAngeles} & Reddit UAE/COVID subreddits \\
Top conf./lift      & 100\% / 2.54 (W1) & 88.8\% / 2.16 (P3/C2) \\
Primary rule value  & Acute coupling & Stable + predictive temporal structure \\
\bottomrule
\end{tabularx}
\vspace{-2 em}

\end{table}

\section{Discussion and Policy Implications}

\paragraph{\textbf{Policy translation}.}
Each validated rule is converted into an operational brief specifying: trigger conditions, expected outcome, statistical strength (support, confidence, lift, conviction, leverage), signal window, agency-specific action playbook, and false-alarm guidance.

\paragraph{\textbf{Policy insights}.}
The rules reveal four actionable crisis-management signals. First, hidden risk can appear under apparent calm: when \feat{health\_concern} rises during \feat{calm\_mobile\_baseline} periods, same-day grocery spikes become more likely (C1, conf.\ 77.4\%), suggesting that NCEMA-style emergency authorities should pair reassurance messaging with supply checks before visible mobility disruption appears. Second, lockdown discourse and consumer preparation can precede vaccine-related discourse by one week (P1, conf.\ 85.3\%), giving MoHAP and health communication teams a lead window for targeted outreach and counter-fatigue messaging. Third, positive sentiment and calm mobility can precede grocery spikes within 48 hours (P3, conf.\ 88.8\%), indicating that recovery narratives should be accompanied by logistics and inventory readiness. Fourth, in the wildfire case, traffic congestion and negative emotion co-occur with governance accountability discourse (W2--W3), suggesting that LAFD, CAL FIRE, and LA County OES should prioritize transparent updates when traffic stress and emotional distress rise together.


\section{Conclusion and Future Work}
This paper introduced an interpretable framework for integrating mobility and social discourse signals to analyze crisis behavior. Across a 33-day LA wildfire prototype and a 671-day UAE COVID-19 case study, the framework extracted cross-domain behavioral structures, validated them through pruning and chronological holdout testing, and translated them into operational guidance. The results show that multimodal behavioral rules provide same-day and short-horizon signals, including 88\% holdout stability and 2--7 day predictive lead windows, supporting communication strategy, logistics readiness, and crisis-response planning. Future work will extend the framework to finer geographic resolution, adaptive dashboards, crisis-calibrated sentiment models, and broader transfer tests across earthquakes, heatwaves, political crises, and other emergencies.
%
\bibliographystyle{splncs04}
\bibliography{refs}

\end{document}